\documentclass{article}
\usepackage[preprint]{spconf}
\usepackage{amsmath,epsfig}
\usepackage{epstopdf}
\usepackage{tabularx}
\usepackage{multirow}
\usepackage{graphicx}
\usepackage{textcomp}
\usepackage{url}
\usepackage{setspace}

\renewenvironment{thebibliography}[1]{%
   \begin{oldthebibliography}{#1}%
     \setlength{\itemsep}{-.3ex}%
}%
{%
   \end{oldthebibliography}%
}

\copyrightnotice{\copyright\ IEEE 2016}
\toappear{To appear in {\it Proc.\ ICME,
                    July 11-16, 2016, Seattle, USA}}

\begin{document}\sloppy
\def\x{{\mathbf x}}
\def\L{{\cal L}}

\title{Weakly Supervised Scalable Audio Content Analysis}
%
\name{Anurag Kumar, Bhiksha Raj}
\address{Language Technologies Institute\\
Carnegie Mellon University, Pittsburgh, PA - USA\\
alnu@andrew.cmu.edu, bhiksha@cs.cmu.edu}
%
%

\maketitle
\begin{abstract}
Audio Event Detection is an important task for content analysis of multimedia data. Most of the current works on detection of audio events is driven through supervised learning approaches. We propose a weakly supervised learning framework which can make use of the tremendous amount of web multimedia data with significantly reduced annotation effort and expense. Specifically, we use several multiple instance learning algorithms to show that audio event detection through weak labels is feasible. We also propose a novel scalable multiple instance learning algorithm and show that its competitive with other multiple instance learning algorithms for audio event detection tasks. 
\end{abstract}
\vspace{-0.20in}
\section{Introduction}
\label{sec:intro}
\vspace{-0.15in}
Multimedia content analysis is a necessity for meaningful retrieval and indexing of multimedia data.  It is becoming even more important due to the explosive growth of multimedia data over the internet. The focus of this paper is the audio component of multimedia which carries a significant amount of information about the overall content of multimedia data. Audio content analysis for us means successful detection of different \emph{acoustic events} in a given audio recording. Audio content analysis or audio event detection (AED), the term more frequently used in this paper is important for reasons other than multimedia retrieval as well. Several applications such as surveillance \cite{1} and monitoring are much easier to do through audio data only. Audio signals are not only easier to capture and transmit but can also pass through obstacles where cameras would be simply useless. An excellent example for audio based monitoring systems are those used in wildlife monitoring,  for example bird species recognition using birdsong audio \cite{ruizmultiple} \cite{stowell2014}.  Moreover, audio event detection plays an important role for context aware systems \cite{eronen2006}.  

A variety of methods including those already referred have been proposed for AED tasks. Taking cues from automatic speech recognition GMM-HMM architectures have been employed for AED as well\cite{zhuang2010}. Discriminative learning methods in which fixed length representations for audio segments combined with discriminative classifiers such as Support Vector Machine or Random Forest classifiers have also been proposed \cite{7} \cite{supbow} \cite{12}\cite{autoenco}. With success of Deep neural networks for ASR, audio event detection using DNNs were attempted in \cite{Ashraf2015} \cite{autoenco} \cite{gencoglu}. An inherent problem with AED is presence of overlapping events. Successful detection of overlapping events is crucial for real world applications. Techniques such as matrix factorization and deep neural networks have been used for this more challenging tasks \cite{dessein2013}\cite{cakir2015}. 

From review of literature on audio event detection it is noticeable that AED is currently mainly driven by supervised learning approaches. The data consists of either examples of audio events or recordings for which time stamps of occurrence of each audio event is provided. We will refer to datasets in these forms as \emph{Strongly} labeled data. Strongly labeled data can be easily used for training a variety of supervised learning algorithms for AED tasks.  

However, one can quickly realize that fully supervised approaches where we need strongly labeled data will run into scaling issues. These scaling issues will be on several fronts. \emph{First}, in terms of amount of training data available for each acoustic event. Creating strongly labeled data requires manually annotating data where all times of occurrences of an event are properly marked for a given recording. Clearly, this is an extremely time consuming and expensive task to perform. Thus generating a few hours of training data (positive examples) for a particular audio event is very difficult. \emph{Second}, in terms of audio events vocabulary for which detectors are built. The number of audio events for which one can create strongly labeled data cannot again be made very large due to the previous reason. In fact to the best of our knowledge there is no large scale strongly labeled dataset meant for audio event detection consisting of more than one hundred audio event in the vocabulary. This obviously puts some constraints on audio event detection research. 

A way out of the above problems is learning through weakly labeled data. \emph{Weak} labels implies information about the presence or absence of an event in the recording is known but the exact time frames where it is present need not be known. If we can learn audio event detection from \emph{weakly} labeled data the scaling issues stated previously can be tackled. The is because one can consider mining the vast amount of data available on the web. For audio/video recordings available on the web \emph{weak} labels can be obtained using the  metadata  (\emph{tags}, \emph{titles} etc.) associated with the recordings. Even if no such metadata is available it is much easier to traverse through a recording and simply mark down if an event is present or not rather than putting exact time stamps of occurrences of the event. This essentially solves both of the problems mentioned in previous paragraph. Surprisingly, weakly supervised AED has received next to negligible attention. A particularly interesting and related work is classification of bird species \cite{ruizmultiple} using weakly labeled birdsong audio through Multiple Instance Learning (MIL). In another work, MIL was used for music genre classification using weak labels \cite{tmmmusic}. 

We intend to show in this work that generic audio event detection using weakly labeled data is feasible through the well known multiple instance learning framework. Multiple Instance Learning (MIL) is a \emph{weakly supervised} learning form in which labels are available for groups or bags of instances rather than each instance itself as is the case in supervised learning. We use several well known multiple instance learning (MIL) algorithms to demonstrate that weakly labeled data can indeed be used for AED tasks. Moreover, we also focus on the scalability issue at the next level; which is the scalability of MIL algorithms themselves for AED tasks. We propose a novel scalable multiple instance learning which is based on ideas similar to another scalable MIL algorithm \cite{wei2014}. Its performance is superior to $2$ of the known MIL algorithm used in this work and offers advantages over the third method in terms of scalability while retaining the performance. 

The rest of the paper is organized as follows. In Section 2 we formulate weakly supervised AED using MIL. In Section 3 we provide description of various MIL algorithms and our proposed MIL algorithm. Section 4 shows experimental results and we conclude in Section 5. 
\vspace{-0.25in}
\section{Multiple Instance Learning and AED}
\vspace{-0.15in}
In supervised learning labels for all instances are known during training and the learner tries to find decision boundaries which can separate these labeled instances in the best possible way. Since we are concerned with detection of presence or absence of an event in a recording we will be basically concerned with binary case where labels are either $-1$ or $1$. The instances belonging to the target class will be referred to as positive ($+1$ label) and those not belonging to target class as negative ($-1$). Multiple instance learning can be described as a generalized form of supervised learning. Here labels are not known for each instance, instead labels are known for a \emph{bag} of instances. A \emph{bag} is labeled as positive if \emph{atleast} one instance in it is known to be positive otherwise it is labeled negative. This implies that all instances in a negative bag can be marked as negative examples for the target class whereas the same cannot be done for a positive bag in which an instance can be positive as well as negative. This imbalance creates a unique form of learning where learner needs to learn decision boundaries using this bag level representations. It means that atleast one of the instance in every positive bag must lie on the positive side of decision boundary. Starting from its utility in drug activity detection\cite{24}, several algorithms such as Diverse Density,  MILES, Citation-KNN, MIBoosting, MiGraph, SVM based approaches etc. have been proposed to learn in this paradigm \cite{amores2013}\cite{doran2014}.  MIL has found applications in several areas such as image classification and retrieval, text categorization, web mining etc. \cite{doran2014} \cite{amores2013} give excellent surveys of theory and applications of MIL. We propose to use it for audio event detection using weakly labeled data. 

Let $\{(X_1,Y_1),...(X_i,Y_i)...(X_M,Y_M)\}$ be the set of $m$ bag-label pairs. $X_i$ represents a bag and $Y_i \in \{-1,1\}$ its label. The instances in the bag $X_i$ are represented by $\mathbf{x}_{ij}$ where $j = 1 \,\,to\,\,m_i$, $m_i$ being the number of instances in the bag $X_i$. Each $\mathbf{x}_{ij}$ is $D$ dimensional vector representing the instance. The instance labels $y_{ij}$ for $x_{ij}$ can be deterministically inferred from $Y_i$ for negative bags only. 
  
Moving specifically into audio, consider a weakly labeled data in which the presence or absence of an event in an audio recording is known. We break this recording into several small contiguous segments. Considering reasonably sized segments, it is safe to assume that if an event was marked to be present in the whole recording then atleast one of the segment is a positive example for that event. One the other hand, if an event is not marked to be present in a recording, clearly, none of the segments will contain that event and hence all segments are negative examples for the event concerned. The whole picture naturally falls into MIL framework. The recordings can be treated as bags $X_i$ and the segments of the recordings as instances of the corresponding bag. From the arguments just stated if the weak information identifies presence of an event in a recording then the label for the corresponding bag is $+1$ for that event otherwise $-1$. Thus, AED using weak annotations can be cast as an MIL problem and one can employ any of the whole well-known MIL algorithm. Before going into the details of the MIL algorithms we need the feature representation for the audio segments in each bag. We use soft-count bag of audio word representation which is described in next Section.
\vspace{-0.15in}
\subsection{Audio Feature Representation}
\label{sec:featrep}
\vspace{-0.1in}
Bag of audio words is a simple yet effect way of representing acoustic events for detection and classification purposes \cite{7}\cite{autoenco}. The first step is to parameterize all audio recordings by basic features such as Mel-Cepstra Coefficients (MFCC). The MFCC features for all the available audios are then used to learn a codebook through clustering algorithms (e.g K-means). Then for any given recording its MFCC vectors are quantized to one of the cluster in the codebook. Finally, a histogram representing the count of number of MFCC vectors belonging to each cluster is used as feature representation for the recording. 

We need a form of feature representation which can robustly represent short duration audio segments (instances of the bags). \cite{shortduration} showed that  it is better to use \emph{soft count} histogram obtained using Gaussian Mixture Model (GMM) for event detection in short audio segments. Hence, we use a GMM based bag of audio words representation. The first step in this case is to learn a universal GMM using the MFCC features of training audio data. Let $\mathcal{G} = \{w_k,N(\vec{\mu}_k, \Sigma_k), k = 1 \,\,to \,\,M\}$ be the obtained GMM, where $w_k$, $\vec{\mu}_k$ and $\Sigma_k$ are the mixture weight, mean and covariance parameters of the $k^{th}$ Gaussian in $\mathcal{G}$. Let the MFCC vectors of an audio with total of $T$ MFCC vectors be represented by $\vec{m}_t, t= 1\,\,to\,\,T$. Then \emph{soft-count} for $k^{th}$ component is computed as 
\begin{align}
Pr(k | \vec{m_{t}}) = & \frac{w_{k}N(\vec{m_{t}} ; \vec{\mu}_k, \Sigma_k)}{\sum\limits_{j=1}^M w_jN(\vec{m_{t}} ; \vec{\mu}_k, \Sigma_k)}\\
P(k) = & \frac{1}{T}\sum\limits_{i=1}^T Pr(k | \vec{m_{t}})
\end{align}
The feature representation of the audio segment is the vector $\vec{P}_M=[P(1),..P(k)..P(M)]^T$. $\vec{P}_M$ is normalized to sum to 1. Thus each instance vector in the bag $\mathbf{x}_{ij}$ is an $M$ dimensional \emph{soft-count} histogram.
\vspace{-0.15in}
\section{MIL Algorithms}
\vspace{-0.15in}
As stated previously, a variety of MIL algorithms have been proposed. Overall, we use three known algorithms for MIL and propose one novel algorithm for multiple instance learning. The first two algorithm referred as miSVM and MISVM \cite{misvm} are based on Support Vector Machine (SVM). The standard SVM algorithm is modified to work in MIL domain. Although a few other formulations of SVM for MIL domain have been proposed \cite{doran2014}, miSVM and MISVM were the first set of SVM formulations for MIL and performs well on a variety of MIL tasks. Our third MIL algorithm is a more generic approach called miFV which tries to address the scalability issue in MIL paradigm. Our proposed novel MIL method is also based on the idea similar to the one behind miFV. We provide a brief description of these methods.
\vspace{-0.15in}
\subsection{miSVM and MISVM}
\vspace{-0.10in}
\textbf{miSVM} is the first method proposed in \cite{misvm} to directly solve MIL using SVM. miSVM is an instance level approach in which the labels for instances in positive bags are treated as unobserved integer variables. It then tries to solve standard SVM by putting in an additional constraint which ensures that the atleast one instance in the positive bag is positive. This formulation is shown in Eq \ref{eq:misvm}. It leads to a mixed integer programming problem where margin maximization is done jointly over labels of instances and the separating hyperplanes. An optimization heuristics is used to solve it in an iterative manner. First, standard SVM is solved using the current assigned labels for instances in positive bags. Second, using the obtained SVM solution labels are assigned to instances in positive bags. All instances in positive bags are initialized with positive label to obtain the first SVM solution. The process is repeated till no change in labels of instances are observed. To ensure positive bag constraint, if all instances in a positive bag gets classified as negative at any step then the instance with maximum output is assigned positive label.
\begin{align}
\label{eq:misvm}
\begin{split}
& \min_{\{y_{ij\}},\textbf{w},b,\xi} \frac{1}{2}||w||^2 + C\sum\limits_{ij}\xi_{ij} \,\,\,\,\, s.t \,\,\, \forall \,\,\, i,j\\
& \,\, y_{ij} (\langle \textbf{w} ,\textbf{x}_{ij} \rangle + b )  \ge 1-\xi_{ij}, \,\, \xi_{ij}\,\ge 0, \,\, y_{ij} \in \{-1,1\} \\
& \sum\limits_{j=1}^{m_{i}} \frac{y_{ij}+1}{2} \ge 1 \,\, \forall Y_i = 1 \,\, \& \,\, y_{ij}=-1 \,\,\forall \,\, Y_i = -1
\end{split}
\end{align}
\textbf{MISVM} is the second approach proposed alongside miSVM. This is a bag level approach in which the goal is to describe margin with respect to bags and then try to directly maximize this margin. The formulation is standard SVM with max constraint as shown in Eq \ref{eq:MISVM}.
\begin{align}
\label{eq:MISVM}
\begin{split}
& \min_{\textbf{w},b,\xi} \frac{1}{2}||w||^2 + C\sum\limits_{i}\xi_{ij} \\
& s.t \,\,\, Y_{i}(\underset{\mathbf{x}_{ij} \in X_i}{max}(\langle \textbf{w} ,\textbf{x}_{ij} \rangle + b ))  \ge 1-\xi_{i}, \,\, \xi_{i}\,\ge 0\\
\end{split}
\end{align}
Essentially, each positive bag is represented by one ``witness'' instance which determines the margin. The optimization heuristic similar to miSVM is used to solve the problem. The only difference is that in this case a selector variable which determines which instance of a positive bag is ``witness'' is updated in each iteration. Both of these methods are valid for non-linear kernel SVMs as well. More details can be found in the original paper \cite{misvm}. 
\vspace{-0.20in}
\subsection{Scalable Algorithms}
\vspace{-0.10in}
Most of the MIL algorithms including miSVM and MISVM suffers from scalability issue \cite{wei2014}. Scalability is an important factor especially when we are trying to work on problems such as AED where total number of instances will become very large even for a few hours of audio data. A scalable multiple instance learning algorithm called miFV was proposed in \cite{wei2014}. The main idea is that instead of trying to learn complex hypothesis for bag representation,  map each bag into a single vector representation, usually in some higher dimensional space. This vector representation for each bag should try to encode as much non-redundant information as possible for the bag. Also, it should be computationally efficient for large scale learning. Once this mapping can be done efficiently the MIL problem effectively moves to the supervised learning paradigm because each bag is represented by a single vector and has a corresponding label. The scalability issue can be further addressed by using any scalable supervised learning algorithm. For example, linear SVMs are known to be computationally cheap and performs quite well in high dimensional space. 
\vspace{-0.15in}
\subsubsection{\textbf{miFV}}
\vspace{-0.10in}
miFV uses Fisher Vectors (FV) for encoding bags into vector representation. FV originates from Fisher Kernels and is a state of art method for image retrieval and classification. We request the reader to refer to \cite{sanchez2013} for a detailed description of Fisher Kernels and Fisher Vectors. \cite{wei2014} claimed that under the assumption of independence for instances in the bags, it can also be used for encoding the bags in MIL framework. Here, we show an outline of obtaining Fisher Vectors for each bag. A Gaussian Mixture Model (GMM) is first used to model instances across all bags. Please note that this different from the one used for audio feature representation. This GMM is learned over the instance space using all instances in training bags. Let this $K$ component GMM be represented as $\mathcal{M} = \{w_k,N(\pmb{\mu}_k, \Sigma_k)\, k = 1 \,\,to \,\,K\}$ where $w_k$ is the mixture weight, $\pmb{\mu}_k$ is mean vector and $\Sigma_k$ is the covariance matrix of $k^{th}$ Gaussian. The Gaussians are assumed to have diagonal covariance with diagonal variance vector represented as $\pmb{\sigma}^2_k$. Then for given a bag $X_i$ with $m_i$ instances we compute the following for each component of GMM
\begin{align}
\gamma_j(k) & = \frac{w_{k}N(\mathbf{x}_{ij}; \pmb{\mu}_k, \Sigma_k)}{\sum\limits_{j=1}^K w_jN(\mathbf{x}_{ij}; \pmb{\mu}_k, \Sigma_k)} \label{eq:gmx}\\
\alpha^{X_i}_{w_k} & = \frac{1}{m_i\sqrt{w_k}} \sum_{j=1}^{m_i} (\gamma_j(k) - w_k) \label{eq:fvmxwt}\\
\alpha^{X_i}_{\pmb{\mu}_k} & = \frac{1}{m_i\sqrt{w_k}} \sum_{j=1}^{m_i} \gamma_j(k) \left( \frac{\mathbf{x}_{ij} - \pmb{\mu}_{k}}{\pmb{\sigma}_{k}} \right) \label{eq:fvmu} \\
\alpha^{X_i}_{\pmb{\sigma}_k} & = \frac{1}{m_i\sqrt{2 w_k}} \sum_{j=1}^{m_i} \gamma_j(k) \left( \frac{(\mathbf{x}_{ij} - \pmb{\mu}_{k})^2}{\pmb{\sigma}^2_{k}} - 1\right) \label{eq:fmsgm}
\end{align}
Finally, the Fisher Vector is concatenation of $\alpha^{X_i}_{w_k}$, $\alpha^{X_i}_{\pmb{\mu}_k}$, $\alpha^{X_i}_{\pmb{\sigma}_k}$ for all $K$ Gaussians. This results in a $(2D + 1)K$ dimensional representation where $D$ is the dimension of instance space. An improved fisher vector (IFV) obtained by sign - square rooting and $L_2$ normalization of the original Fisher vector leads to superior performance \cite{sanchez2013}. Once fisher vector has been used one can potentially use any supervised learning method to obtain audio event detectors. In patricular, Fisher vectors work remarkably well with linear SVMs. Its clear from Equations \ref{eq:fmsgm}, \ref{eq:fvmu} and \ref{eq:fvmxwt} that once the GMM $\mathcal{M}$ has been obtained, the computation of Fisher vector is cheap and can be done efficiently. Combined with linear SVMs it can efficiently address the scalability issue in MIL. 
\vspace{-0.20in}
\subsubsection{\textbf{Proposed MIL Algorithm}}
\vspace{-0.1in}
Here we propose another scalable multiple instance learning algorithm. The central idea is same as the one behind miFV which is to map bags to single vector representation. We propose an alternative and efficient method of mapping bags to single vectors. Similar to miFV we also start with the same GMM $\mathcal{M}$ trained in the instance space. The single vector representation for a bag is then obtained by maximum a posteriori (MAP) adaptation of instances in the bag to the GMM $\mathcal{M}$. This essentially implies that the parameters of $\mathcal{M}$ are updated to effectively represent the instances in the bag. The encoding of a bag $X_i$ is done by following steps. 

For each component of the GMM we compute the posterior probabilities $\gamma_j(k)$ as in Eq \ref{eq:gmx}. Then, the mean and variance updates are computed as
\begin{align}
\beta^{X_i}_{\pmb{\mu}_k} & = \frac{\sum_{j=1}^{m_i} \gamma_j(k) \,\, \mathbf{x}_{ij} + r \pmb{\mu}_{k}}{\sum_{j=1}^{m_i} \gamma_j(k) + r} \label{eq:spmu}\\
\beta^{X_i}_{\pmb{\sigma}_k} & = \frac{\sum_{j=1}^{m_i} \gamma_j(k) \,\, \mathbf{x}_{ij}^2 + r (\pmb{\mu}^2_{k}+  \pmb{\sigma}^2_{k})} {\sum_{j=1}^{m_i} \gamma_j(k) + r} - (\beta^{X_i}_{\pmb{\mu}_k})^2 \label{eq:spsgm}
\end{align}
These updates can be derived using the general MAP estimation equations \cite{bimbot}\cite{gauvain1994} . The factor $r$ controls the amount by which the parameters ($\pmb{\mu}_{k}$ , $\pmb{\sigma}_k$) from $\mathcal{M}$ affect the new estimates $\beta^{X_i}_{\pmb{\mu}_k}$ and $\beta^{X_i}_{\pmb{\sigma}_k}$. Although there is a known update form for mixture weights, it does not directly come from MAP estimation and we do not use employ mixture weight updates in our work. In fact the authors of miFV uses VLFeat  \cite{vlfeat} for Fisher Vector implementation which also considers only mean (Eq \ref{eq:fvmu}) and variance (Eq \ref{eq:fmsgm}) parts. The Fisher Vectors are thus $2KD$ dimensional. We concatenate $\beta^{X_i}_{\pmb{\mu}_k}$ and $\beta^{X_i}_{\pmb{\sigma}_k}$ for all GMM components to obtain the $2KD$ dimensional extended vector representation for the bag $X_i$.   

GMM-MAP adaptation is a state of art method for speaker identification \cite{bimbot} \cite{campbell2006} and the concatenated vectors over all $K$ components are referred as Supervector. Hence, we name our proposed method for multiple instance learning as miSUP. As stated before, we can use both mean (Eq \ref{eq:spmu}) and variance (Eq \ref{eq:spsgm}) vectors in our Supervector representation for bags. However, as we show empirically later that the concatenation of only means updates (Eq \ref{eq:spmu}) are sufficient to obtain reasonably good results. This is an important aspect of miSUP because the dimensionality of the bag representation will be now only \emph{KD} instead of \emph{2KD} in miFV. This reduction by half can be significant if \emph{KD} is large and can speed up the learning process in the next stage where classifiers are trained. It can also save a significant amount of storage space if we go into really large scale audio (multimedia) content analysis problems where feature vectors of possibly millions or billions of recordings (bags) are to be stored. We will refer to the mean only Supervector based miSUP method as miSUP\textunderscore MN. Overall, we will be looking into a total of $5$ MIL (miSVM, MISVM, miFV, miSUP, miSUP\_MN) based frameworks for weak labels based AED.  
\vspace{-0.15in}
\section{Experiments and Results}
\vspace{-0.10in}
We used TRECVID-MED 2011 database \cite{32} for evaluation of different MIL algorithms. Weak labels are obtained for $8$ acoustic events on a subset of the database. Binary classifiers are learned to detect presence or absence of each of these events. The total number of recordings or bags used in the experiment is $457$. From here on we will refer to this set of recordings as dataset. The total size of the dataset is about 22 hours. The number of positive bags for each event in the dataset is as follows. \emph{Cheering}-171, \emph{Children Voices} - 33, \emph{Clapping} - 102, \emph{Crowd} - 142, \emph{Drums} - 25, \emph{Engine Noise} - 80, \emph{Laughing} - 102 \emph{Scraping} - 30. Please note that a bag can be positive for multiple events and it can also be possibly negative for all $8$ events considered. The dataset is divided into $4$ sets. $3$ sets are used for training the model which is then tested on the fourth set. We do this all $4$ ways (\emph{i.e} each set becomes the test set) so that results on the whole dataset is obtained.  

We use a uniform segmentation scheme on the recordings to obtain the instances for each bag. Each instance is a one second segment of the recording, with segmentation done in an overlapping manner ($50\%$ overlap). The number of instances varies significantly from bag to bag due to variation in the length of the recordings. We use 21-dimensional Mel Frequency Cepstrum Coefficient (MFCC) vectors to represent audio over which \emph{soft-count} representations $\vec{P}_M$ as described in \ref{sec:featrep} are obtained. MFCC vectors are computed over an analysis window of $20ms$ with $10ms$ overlap between adjacent windows. We use $M=64$ and $M=128$ for $\vec{P}_M$. All experiments are performed for both representations. Although, features representing audio segments are important, due to space constraints we would be showing only better of the two results ($M=64$,$M=128$) for all events under a given MIL algorithm. We use linear SVMs (LIBLINEAR \cite{fanliblinear}) in all MIL algorithms.
\begin{figure}[t]
\centering
\includegraphics[width=0.8\columnwidth,height=1.5in]{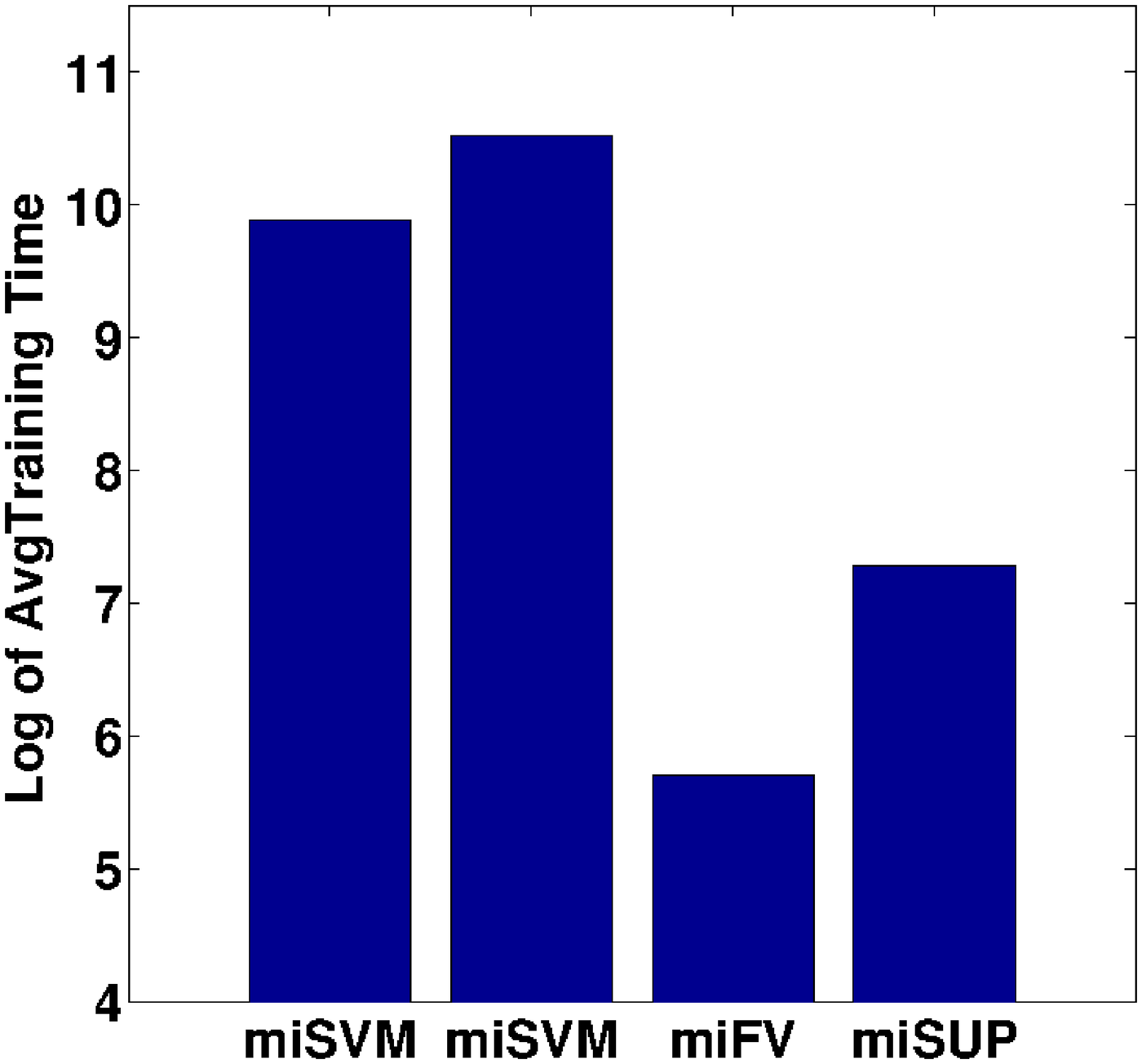}
\caption{Comparison of Mean Training Times (Log(seconds))}
\label{fig:runtime}
\vspace{-0.25in}
\end{figure}

\textbf{Overall Detection Results:} Average Precision (AP) for each event is used as performance metric. The Mean Average Precision over all events is also shown for each case. Table \ref{tab:allres} shows AP numbers for all events and methods. The first important noticeable aspect in Table \ref{tab:allres} is that miFV and miSUP are far more superior compared to miSVM and MISVM. There is an absolute improvement of about $16-17\%$ in MAP value using these methods. For several events miFV and miSUP improves AP by more than $20\%$  over miSVM and MISVM. MISVM seems to be marginally better than miSVM. Also, miFV seems to be performing slightly better than miSUP. We make note of the fact that we used improved Fisher Vectors (IFV) \cite{sanchez2013} in miFV approach. Its possible that some normalization techniques for miSUP might lead to better results. The miSUP$\_$MN method with only half the dimension of miFV and miSUP is competitive with both methods by giving reasonably good performance.

\begin{table}[t]
\centering
\caption{Average Precision for Each Event}
\resizebox{\columnwidth}{!}{
\begin{tabular}{|c|c|c|c|c|c|}
\hline  
Events & miSVM & MISVM & miFV & miSUP & miSUP$\_$MN \\
\hline 
Cheering & 0.482 & 0.500 &0.629&0.605&0.59 \\
\hline
Children Voices & 0.130 & 0.142 & 0.264&0.193&0.193\\
\hline
Clapping & 0.383 & 0.395 & 0.492&0.494&0.477 \\
\hline
Crowd & 0.470 & 0.583 & 0.685&0.670&0.663\\
\hline
Drums & 0.078 & 0.112 & 0.233&0.263&0.242\\
\hline$  $
Engine Noise & 0.330 & 0.389 & 0.608&0.583&0.540 \\
\hline
Laughing & 0.288 & 0.300 & 0.506&0.431&0.425 \\
\hline
Scraping & 0.449  & 0.305 & 0.562&0.539&0.511\\
\hline
\textbf{MAP} & \textbf{0.328} & \textbf{0.339} & \textbf{0.497} & \textbf{0.472}& \textbf{0.455}\\
\hline
\end{tabular}
}
\label{tab:allres}
\vspace{-0.15in}
\end{table}
An important parameter in miFV and miSUP approach is the component size $K$ of  GMM ($\mathcal{M}$) learned in instance space. It determines the overall dimensionality of vector representing bags and can affect the learning in a major way. We  study this aspect for 7 different values  $1,4,8,16,32,64,128$. Table \ref{tab:compres} shows the AP values corresponding to each of these values for miFV and miSUP methods. 
\begin{table*}[t]
\label{tab:compres}
\centering
\caption{AP Values For different $K$ in GMM $\mathcal{M}$}
\resizebox{1.8\columnwidth}{!}{
\begin{tabular}{|c|c|c|c|c|c|c|c|c|c|c|c|c|c|c|}
\hline  
K $\longrightarrow$ &\multicolumn{2}{c|}{1} & \multicolumn{2}{c|}{4} &  \multicolumn{2}{c|}{8} & \multicolumn{2}{c|}{16} & \multicolumn{2}{c|}{32} & \multicolumn{2}{c|}{64} & \multicolumn{2}{c|}{128}\\ 
\hline
Events $\downarrow$ &miFV&miSUP&miFV&miSUP&miFV&miSUP&miFV&miSUP&miFV&miSUP&miFV&miSUP&miFV&miSUP\\
\hline 
Cheering & 0.621&0.574&0.629&0.566&0.629&0.579&0.603&0.605&0.61&0.58&0.598&0.582&0.624&0.574\\
\hline
Children Voices & 0.264&0.171&0.227&0.182&0.199&0.158&0.174&0.193&0.176&0.185&0.152&0.142&0.172&0.125\\
\hline
Clapping & 0.492&0.445&0.472&0.471&0.457&0.474&0.461&0.494&0.441&0.449&0.435&0.483&0.45&0.445\\
\hline
Crowd & 0.671&0.666&0.664&0.644&0.685&0.646&0.676&0.67&0.674&0.599&0.647&0.601&0.643&0.613\\
\hline
Drums & 0.134&0.263&0.207&0.205&0.233&0.144&0.152&0.181&0.217&0.126&0.197&0.19&0.204&0.106\\
\hline
Engine Noise & 0.582&0.583&0.608&0.535&0.55&0.502&0.572&0.537&0.579&0.53&0.592&0.513&0.577&0.486\\
\hline
Laughing & 0.429&0.38&0.465&0.403&0.506&0.41&0.479&0.406&0.487&0.431&0.471&0.413&0.468&0.416\\
\hline
Scraping & 0.521&0.434&0.547&0.356&0.562&0.532&0.546&0.498&0.458&0.284&0.537&0.539&0.546&0.444\\
\hline
MAP & 0.464&0.439&0.477&0.420&0.478&0.430&0.457&0.448&0.455&0.398&0.454&0.433&0.461&0.401\\
\hline
\end{tabular}
}
\vspace{-0.15in}
\end{table*}

\textbf{Analysis of Algorithms}:
We try to analyze the average training time for each MIL algorithm. For a given MIL algorithm the average training times for each event is noted and then a mean training time is obtained by averaging over all events. This comparison is shown in Figure \ref{fig:runtime}. The y-axis shows log of the mean training times in seconds. Clearly, miSUP and miFV are much faster (about $20\,\,to\,\,100$ times) compared to miSVM and MISVM. The order of difference in time is actually higher for several individual events. This demonstrates the higher scalability of miFV and miSUP compared to miSVM and MISVM. The comparison between miFV and miSUP is not completely fair because Fisher Vectors were implemented using \cite{vlfeat} which is possibly a very optimized implementation. With optimized implementation miSUP should be closely comparable to miFV. For classifiers such as linear SVMs the gain obtained by reducing dimension with MISUP$\_$MN does not give substantial reduction in classifier training time and is similar to MISUP\_MN. However, its important to note that for other classifiers this gain can be significant. 
\vspace{-0.25in}
\section{Discussions and Conclusions}
\vspace{-0.15in}
In this paper our goal was to address scalability issues of audio event detection. We addressed scalibility issues on two fronts. \emph{First}, is directly related to audio event detection where obtaining \emph{strongly} labeled data is extremely difficult and expensive. This puts constraints in terms of vocabulary of audio events and amount of training data available for audio events. The goal of this work was to show a way where these scaling issues can be addressed. Our  assertion is that obtaining weak labels is much easier compared to strong labels. We then, successfully showed that multiple instance learning framework can be used to learn from weakly labeled data. The \emph{second} front is the scalability of MIL algorithm itself for AED tasks. Standard MIL algorithms such as miSVM and MISVM can run into scalability issues because number of instances will run into hundreds of thousands even for medium sized audio dataset. Hence, we considered algorithms which are specifically meant to address scalability in multiple instance learning. We proposed a novel MIL algorithm and showed that it is competitive with the other scalable MIL algorithm existing in literature. Interestingly, for audio event detection these scalable MIL algorithms turned out to be far more superior compared to standard MIL algorithms such as miSVM and MISVM which are known to perform well in general. Overall miFV performed best and miSUP and miSUP$\_$MN came close to it. We also showed that MAP adaptation of means only (Eq \ref{eq:spmu}, miSUP$\_$MN) where the dimension of bag vector representation is half of miFV can work reasonably well. They can not only address memory issues due to lower dimensions but can also reduce classifier training time. Even though data size (22 hours) was limited, our overall approach and good performance on this dataset does throw some light on scalable audio content analysis. 
\vspace{-0.15in}
\bibliographystyle{IEEEbib}
\ninept
\bibliography{references}
\end{document}